**The potential energy landscape and inherent dynamics of a hard-sphere fluid**

Qingqing Ma and Richard M. Stratt

Department of Chemistry

Brown University

Providence, RI 02912

# ABSTRACT

Hard-sphere models exhibit many of the same kinds of supercooled-liquid behavior as more realistic models of liquids, but the highly non-analytic character of their potentials makes it a challenge to think of that behavior in potential-energy-landscape terms. We show here that it is possible to calculate an important topological property of hard-sphere landscapes, the geodesic pathways through those landscapes, and to do so without artificially coarse-graining or softening the potential. We show, moreover, that the rapid growth of the lengths of those pathways with increasing packing fraction quantitatively predicts the precipitous decline in diffusion constants in a glass-forming hard-sphere mixture model. The geodesic paths themselves can be considered as defining the intrinsic dynamics of hard spheres, so it is also revealing to find that they (and therefore the features of the underlying potential-energy landscape) correctly predict the occurrence of dynamic heterogeneity and non-zero values of the non-Gaussian parameter. The success of these landscape predictions for the dynamics of such a singular model emphasizes that there is more to potential energy landscapes than is revealed by looking at the minima and saddle points.

# I. INTRODUCTION

The key to understanding the structure of liquids turned out to be the simple, but profound, realization that the geometrical arrangements of at least simple liquids are determined almost entirely by the need to pack hard objects in the most random way possible.[1,2] That meant, in turn, that the hard-sphere fluid was the obvious starting point for understanding the essentials behind the equilibrium properties of ordinary liquids.[3] But is the same true of the dynamical properties,[4-6] and, in particular, of the unusual dynamics seen in supercooled liquids?[7,8] Even if one limits oneself to fragile examples such as *o*-terphenyl,[9,10] which, under more prosaic conditions ought to be a textbook simple liquid, do we expect the dynamics of a hard-sphere fluid to help us understand the precipitous slowing down that occurs as one begins to approach the glass transition?

Hard spheres actually make an attractive model for studying this and other phenomena characteristic of supercooled liquids.[11-24] Not only are (near) hard-sphere systems mimicked experimentally by colloidal suspensions,[25-29] but dense hard-sphere-mixture simulations exhibit many of the same types of puzzling behavior as their low-temperature soft-potential counterparts. Hard spheres display dynamical heterogeneity,[15,16,27,30-32] non-Gaussian displacement distributions,[16,27,30,31,33] and a wealth of time-scale-dependent behaviors.[34] The question, from our perspective is how this set of phenomena arises in such a simple system.

This paper examines this issue by investigating the potential-energy landscape of a hard-sphere fluid.[35,36] We begin by first noting that there actually is something one can call a fully microscopic potential energy landscape for hard-sphere systems and then showing that we can both investigate it and compare it to the analogous landscapes of soft-particle systems. We should emphasize at the outset that for us a "microscopically-defined potential energy landscape" means a complete, and exact, specification of the potential energy of our system $V$ as a function of the set of individual atomic coordinates $\mathbf{R} = (\mathbf{r}_1, \ldots, \mathbf{r}_N)$. Since V is either 0 or $\infty$ for hard spheres, our function is obviously highly singular, but we explicitly wish to avoid either replacing the hard-sphere potentials with softer potentials [23,37-40] or replacing the potential energy with a free energy [32,41-49] using some sort of coarse-graining scheme. Both approaches have been employed in the literature to good effect, but neither seems particularly well suited to asking if hard spheres *per se* are a good zeroth-order model for slow dynamics.

The other principal issue to confront is what we mean by "investigating" a potential energy landscape. It has become conventional to associate landscape perspectives on dynamics with the properties (and interconnectedness) of the inherent structures of the landscape – the local minima of the function $V(\mathbf{R})$.[50-55] True hard-spheres, though, have no inherent structures of this sort (except in the trivial sense that every allowed configuration is technically an energy minimum and therefore an inherent structure). An intriguing alternative that does apply to both hard and soft systems is to redefine inherent structures as the local minima of the system's *volume*.[56-58] That makes the inherent structures locally jammed packings of the spheres [11,14,35,37,59-61] – a physically appealing result that offers the promise of a landscape picture of the phenomena associated with jamming.[2,62] However, it is not entirely clear what the relationship is between a potential-energy-defined landscape and a volume-defined landscape, nor is it clear what implications specific features of volume landscapes have for dynamics. Given that we are trying to compare landscape perspectives on soft-particle and hard-sphere dynamics, it would seem to behoove us to try to see if there is more to potential-energy landscapes than simply their inherent structures.

The alternative that we employ here is the *landscape-geodesic* approach,[63-65] a strategy that has proven useful in a number of liquid-state applications.[64,66,67] The question posed now is not what the optimum configurations in the landscape are, but what the optimum pathways are that travel from configuration to configuration within our equilibrium ensemble – an idea that is equally well defined for hard and soft potentials.[68] The importance of optimum pathways can be appreciated most easily by noting that the Green's function for diffusing in a time $t$ from some configuration $\mathbf{R}_i$ to some configuration $\mathbf{R}_f$ in the absence of any forces, (but subject to what may be complicated boundary conditions) is given by a path integral over all of the possible paths $\mathbf{R}(\tau)$ obeying the desired boundary conditions ($0 \leq \tau \leq t$).

$$G\left(\mathbf{R}_i \rightarrow \mathbf{R}_f ; t\right) = \int D\mathbf{R}(\tau) \exp\left\{-\frac{1}{4D} \, S[\mathbf{R}(\tau)]\right\} \quad , \tag{1.1}$$

$$(\mathbf{R}(0) = \mathbf{R}_i, \, \mathbf{R}(t) = \mathbf{R}_f) \quad ,$$

$$S[\mathbf{R}(\tau)] = \int_0^t d\tau \left(\frac{d\mathbf{R}}{d\tau}\right)^2$$

When the diffusion constant $D$ is small, though, the paths that dominate are those that make the action $S[\mathbf{R}(\tau)]$ as small as possible – the *geodesic* (shortest) paths.[64] The geodesic paths therefore prescribe what are, in some sense, the inherent dynamics of a diffusing system.[69]

Our systems, of course, do have forces operating in them. In our previous applications of these ideas to particles with soft inter-particle potentials, we found that we could make use of Eq. (1.1) simply by working within what we called the potential-energy-landscape ensemble, the set of all particle configurations $\mathbf{R}$ whose potential energy $V(\mathbf{R}) \leq E_L$, with $E_L$ some landscape energy.[63] The landscape ensemble is thermodynamically equivalent to the canonical ensemble, and the landscape energy can be thought of as a thermodynamic variable conceptually equivalent to (and computationally determined by) the temperature $T$. However, in the landscape ensemble all allowed configurations have equal probability densities, so it is effectively an ensemble with no forces; the potential surface does make its effects felt, but only through what may be horribly complicated boundary conditions embodied in the restriction $V(\mathbf{R}) \leq E_L$.[63]

For hard-sphere systems, there is no unique configurational temperature to define a landscape energy. Still, the distribution of configurations in the hard-sphere potential-energy landscape with $E_L = 0$ is, in fact, identical to that prescribed by the canonical ensemble at any finite temperature. So, provided we can find them, geodesic pathways in hard-sphere potential energy landscapes ought to play the same role as those in soft potential systems. The task of finding geodesics for such a singular landscape is one we will confront in this paper, but it is worth remembering that because the potential-energy-landscape ensemble imposes an all-or-nothing restriction on the allowed liquid configurations, soft-potential landscapes actually impose restrictions on geodesic pathways that are no less stringent than those seen with hard spheres.[64]

So, what can we learn if we do succeed? The first piece of information that comes out of a geodesic analysis is a landscape perspective on *why* diffusion ever becomes slow. Since there are no potential energy barriers in the potential-energy-landscape ensemble, diffusion can become very slow only when the geodesic paths become long and convoluted.[70-75] Indeed, by exploiting the analogy between the path-integral treatment of diffusion around a hard obstacle [76] and our own problem of diffusion restricted to lie outside a given contour of the potential energy, we found that diffusion constants should scale with the ratio of the direct distance between the configurations $\Delta R$ to the length of the geodesic path, $g$.

$$D \sim \lim_{\Delta R \to \infty} \left( \frac{\Delta R}{g} \right)^2 \qquad (1.2)$$

From our landscape perspective, then, it is how lowering the landscape energy necessitates taking ever more circuitous routes around the potential energy landscape that accounts for the dramatic slowing down seen in supercooled liquids.[64]

The principal technical message of this paper is that this same set of ideas works quantitatively well for hard sphere systems – and the success of Eq. (1.2) for hard-sphere systems is evidence that a pathway-based landscape view is a powerful way of understanding the onset of glassy dynamics. More to the point, though, the fact that this treatment shows the same kinds of results for both hard and soft potentials reinforces the notion that hard-sphere-like dynamics is an essential piece of supercooled-liquid behavior.

The remainder of this paper will be organized as follows: In Section II, we develop an approach for finding geodesics in hard-sphere fluids. Section III presents some of the computational details we employ in the course of applying our formalism in Sec. IV to a standard literature binary-hard-sphere mixture model. We also illustrate in Sec. IV some of the ways that geodesic analysis helps illuminate the quantitative significance of the non-Gaussian parameter and the onset of dynamical heterogeneity. We conclude in Sec. V with comments on some of the connections between geodesic theory and both facilitated-kinetics and more conventional landscape theories.

## II. FINDING GEODESIC PATHS FOR HARD-SPHERE SYSTEMS

It has proven relatively straightforward to find geodesic pathways in liquids governed by continuous intermolecular potentials. As we described in a number of studies of atomic and molecular liquids,[64,66,67] the first task is invariably to select from the equilibrium ensemble a number of pairs of liquid configurations $\mathbf{R}(i)$ and $\mathbf{R}(f)$ that can serve as initial ($i$) and final ($f$) liquid configurations for geodesics. Each pair has to be separated by a direct distance $\Delta R = |\mathbf{R}(f) - \mathbf{R}(i)|$ large enough to apply Eq. (1.2), but that can usually be accomplished via a standard (canonical ensemble) Monte Carlo simulation or a (microcanonical ensemble) molecular dynamics simulation. Finding the actual geodesic – the minimum-length path between the endpoints obeying $V(\mathbf{R}) \leq E_L$ at each point along the way – is then an exercise in inequality-constrained optimization. The solutions to such problems are shaped by the Kuhn-Tucker theorem,[77] which in our case, requires that geodesic be some combination of direct steps toward the final endpoint and segments lying along the $V(\mathbf{R}) = E_L$ boundary.[64]

For continuous potential, paths consistent with this requirement arise automatically if we continually take small steps in the direction of the final configuration, provided we make liberal use of the gradient of the potential to guide us back towards the boundary of the energy-allowed region whenever we find ourselves taking on too high a potential energy.[64,67] However, hard-core potentials would seem to call for a rather different strategy. We present one such strategy here.

## A. Direct steps

As in the continuous-potential cases, if the liquid's configuration at time step $t$ is given by $\mathbf{R}(t) = (\mathbf{r}_1(t), \ldots, \mathbf{r}_N(t))$ and the desired final configuration is $\mathbf{R}(f) = (\mathbf{r}_1(f), \ldots, \mathbf{r}_N(f))$, then the system always attempts to move directly towards the final configuration if possible.[64]   In practice, these (attempted) moves in the 3N-dimensional configuration space are specified in terms of a total direct-step length, $\delta_{\text{total}}$

$$\mathbf{R}^{trial\ 1}(t+1) = \mathbf{R}(t) + \delta_{\text{total}}\ \hat{\mathbf{R}}_{\text{t,f}} \quad , \tag{2.1}$$

where, for the remainder of the paper, we define unit vectors and magnitudes for any vectors $\mathbf{X}$ (and in any dimensional space) by

$$\hat{X}_{\text{A,B}} = \left(\mathbf{X}_\text{B} - \mathbf{X}_\text{A}\right)\!\big/ X_{AB} \quad , \qquad X_{AB} = \left|\mathbf{X}_\text{B} - \mathbf{X}_\text{A}\right| \quad .$$

For each particle $j = 1, \ldots, N$, that means that

$$\mathbf{r}_j^{trial\ 1}(t+1) = \mathbf{r}_j(t) + \delta_{\text{j,d}}\ \hat{r}_{j(t),j(f)} \quad , \quad \delta_{j,d} = \delta_{total}\!\left(r_{j(t),j(f)}\big/ R_{t,f}\right) \tag{2.2}$$

so that one would expect typical single-particle direct steps sizes $\delta_{\text{j,d}}$ to be of the order of

$$\delta_d = \delta_{total}\big/\sqrt{N} \quad .$$

## B. Collision avoidance steps

If, as frequently the case, direct moves towards the endpoint would lead to overlaps between particles, we need to do something different.  In our previous studies with soft interparticle potentials, we used the impermissible configurations resulting from direct steps (configurations which had a potential energy $V(\mathbf{R})$ greater than the maximally allowed landscape energy $E_L$) as the starting points of *escape steps* in which the system followed the 3N-dimensional gradient of the potential energy down to the nearest configuration on the boundary $V(\mathbf{R}) = E_L$. These steps were, in general, cooperative, multiple-particle moves, but the end result of repeated iterations of this process was the creation of candidate geodesic pathways that stayed close to the most probable equilibrium locations for the system – the system's energy boundaries.[64,78]

A related strategy could be implemented for a hard sphere system by imagining replacing the system's potential energy by a fictitious potential that is identically zero when the particles do not overlap (as with hard spheres), but is finite and positive-definite whenever particles do overlap.[79]  For a mixture of hard sphere species (with the species labeled by α and β), we can do so by writing

$$V(\mathbf{R}) = \sum_{\substack{\alpha,\beta \\ \alpha \neq \beta}} \sum_{j=1}^{N_\alpha} \sum_{k=1}^{N_\beta} u_{\alpha\beta}\left(r_{jk}\right) + \sum_{\alpha} \sum_{\substack{j,k=1 \\ j<k}}^{N_\alpha} u_{\alpha\alpha}\left(r_{jk}\right) \quad , \qquad (2.3)$$

$$u_{\alpha\beta}(r) = \begin{cases} v_{\alpha\beta}\left(r/\sigma_{\alpha\beta}\right) & r < \sigma_{\alpha\beta} \\ 0 & r > \sigma_{\alpha\beta} \end{cases} \quad , \quad v(x) > 0 \text{ for all } 0 \leq x < 1. \quad (2.4)$$

Whenever the landscape energy $E_L = 0$, the allowed configurations of this new system (and therefore its geodesics) are absolutely *identical* to those of the parent $v(x) = \infty$ hard sphere system.  There are therefore no errors or approximations introduced by such a revision, regardless of the choice made for the function $v(x)$, as long as all of the $v(x)$ functions obeys the conditions shown.  If we further insist that all $v'(x) < 0$, $(0 \leq x < 1)$, then configurations in which the hard spheres overlap even have well-defined fictitious forces $\mathbf{f}_j$ telling each sphere $j$ how to disengage from its overlap with neighboring spheres $k$.

$$\mathbf{f}_j^\alpha = -\nabla_{j_\alpha} V(\mathbf{R}) = \sum_{\beta \neq \alpha} \sum_{k=1}^{N_\beta} \mathbf{f}_{jk}^{\alpha\beta} + \sum_{k>j}^{N_\alpha} \mathbf{f}_{jk}^{\alpha\alpha} \quad , \tag{2.5}$$

$$\mathbf{f}_{jk}^{\alpha\beta} = -\nabla_{j_\alpha} u_{\alpha\beta}\left(r_{j_\alpha,k_\beta}\right) = \begin{cases} \frac{1}{\sigma_{\alpha\beta}} v'_{\alpha\beta}\left(r_{jk}/\sigma_{\alpha\beta}\right)\hat{r}_{jk} & r_{jk} < \sigma_{\alpha\beta} \\ 0 & r_{jk} > \sigma_{\alpha\beta} \end{cases} \tag{2.6}$$

With these forces in hand, one could certainly try to apply precisely the same geodesic-path-finding algorithm that we have been using for soft potentials.[64,67]  What one discovers, though, is that systems whose potential-energy landscape is prescribed by Eqs. (2.3) and (2.4) have a basic distinctions from soft-potential systems (regardless of the specific choice of $v$): any violation of the landscape-energy condition is now fundamentally *local*. Unlike the situation with realistic potentials, these finite range, positive-semi-definite, potential systems cannot "share the pain" of a locally unfavorable contribution to the potential energy by means of a compensatory move involving the remainder of the system.  No matter how many distant particles rearrange, these systems can never make up for the overlap of even a single pair of particles.  As a result, following the (negative of the) 3N-dimensional gradient of our fictitious potential $V(\mathbf{R})$ out of the forbidden area would primarily involve separating small clusters of particles.

Does this locality significantly change the character of the geodesic paths from those taken with softer potentials?  That issue is part of what we want to explore here, but, regardless of the answer, the locality offers us a way to redesign our path-finding algorithm.  Instead of having to use an escape step to propel the entire configuration towards the allowed region, we can rescind steps that would have violated the hard-sphere non-overlap condition in the first place; the locality allows us to replace such moves with *collision avoidance steps* separating precisely those particles that would have collided. That is, the steps move those specific particles in the directions that the fictitious forces $\mathbf{f}_j$ would have prescribed had the particles ventured into the forbidden region. In particular, suppose that time step ($t$) is the last successful step (the last step within the allowed region) and suppose that using Eq. (2.1) to predict a trial ($t+1$)-th configuration step would lead to a collision for some particles $j$. Since the direction of the $j$-th component of the negative configuration-space gradient is prescribed by the three-dimensional unit vector corresponding to the $\mathbf{f}_j$ in Eq. (2.5),

$$\hat{g}_j^{(1)} \equiv \hat{f}_j\left(\mathbf{R}^{trial\ 1}(t+1)\right) \tag{2.7}$$

we can avoid the collision by adopting a new trial candidate for the (t+1)-st configuration:

$$\mathbf{r}_j^{trial\ 2}(t+1) = \mathbf{r}_j(t) + \delta_{ca}\ \hat{g}_j^{(1)} \tag{2.8}$$

where $\delta_{ca}$ is some predetermined collision-avoidance step size.[80] This same procedure can be carried out for every particle that Eq. (2.1) would have forced to collide.

If, for example, Eq. (2.1) would have led a single pair of particles, 1 and 2, to collide with each other (and with no other particles), then it is straightforward to show that Eqs. (2.7) and (2.8) would always generate the same shift in particle positions,

$$\mathbf{r}_1^{trial\ 2}(t+1) = \mathbf{r}_1(t) - \delta_{ca}\ \hat{r}_{12}^{trial\ 1}(t+1)$$

$$\mathbf{r}_2^{trial\ 2}(t+1) = \mathbf{r}_2(t) + \delta_{ca}\ \hat{r}_{12}^{trial\ 1}(t+1) \tag{2.9}$$

a fixed-length displacement of the pair in opposite directions along their line of centers, regardless of the specific choice of $v(x)$.

## C. Iterating the collision avoidance

As the density becomes higher, it becomes increasingly likely that attempts at collision avoidance moves will themselves lead to collisions. A separation of particles 1 and 2 might lead to a collision with some particle 3. Consistent with the basic approach we just outlined, such moves are also rescinded, and we repeatedly try to use the forces that would have come into play had the particles overlapped to find a move that would not lead to a collision for any of the particles involved.

This procedure could be implemented in a number of ways, but we have found that the following scheme is reasonably effective. For each successive trial $T$, we define the weighted force vectors for each particle

$$\mathbf{F}_j^{(T)}(t+1) = \left[ c_j \, \hat{f}_j \right]_{\mathbf{R}} trial \; T_{(t+1)} \tag{2.10}$$

$$c_j = \left| \mathbf{f}_j \right| \Big/ \sqrt{\sum_i \left| \mathbf{f}_i \right|^2} \qquad , \qquad \hat{f}_j = \mathbf{f}_j \Big/ \left| \mathbf{f}_j \right| \tag{2.11}$$

and generalize Eqs. (2.7) and (2.8) to read

$$\hat{g}_j^{(m)}(t+1)) = \sum_{T=1}^{m} \mathbf{F}_j^{(T)}(t+1) \Big/ \left| \sum_{T=1}^{m} \mathbf{F}_j^{(T)}(t+1) \right| \tag{2.12}$$

$$\mathbf{r}_j^{trial \; (m+1)}(t+1) = \mathbf{r}_j(t) + \delta_{ca} \, \hat{g}_j^{(m)} \tag{2.13}$$

As the reader can easily verify, for the first iteration ($m=1$), Eqs. (2.12) and (2.13) are identical to Eqs. (2.7) and (2.8). In particular, the only dependence on the magnitudes (as opposed to the directions) of the fictitious forces, that embedded in the coefficients $c_j$ defined by Eq. (2.11), completely cancels at this stage.  Once the iteration continues, these coefficients allow the algorithm to use the relative magnitudes of the forces to apportion the collision avoidance moves among the particles that need it the most in each iteration.  However, even then, all of the specifics regarding the force magnitudes continue to cancel out whenever the collisions happen in pair-wise sequences.

We can illustrate how this process occurs with the example we mentioned earlier: If an attempted direct move were to lead to particle 1 colliding with particle 2, and if the first-iteration collision avoidance move would cause particle 2 to collide with particle 3, the two-particles-interacting-at-a-time character of the moves would yield coefficient values $c_1 = c_2 = 1/\sqrt{2}$ , $c_3 = 0$ in the first trial and $c_1 = 0$ , $c_2 = c_3 = 1/\sqrt{2}$ in the second. Hence the trial particle positions after the second iteration of the collision avoidance (the successor to Eq. (2.9)) would be

$$\mathbf{r}_1^{trial\ 3}(t+1) = \mathbf{r}_1(t) - \delta_{ca}\ \hat{r}_{12}^{trial\ 1}(t+1)$$

$$\mathbf{r}_2^{trial\ 3}(t+1) = \mathbf{r}_2(t) + \delta_{ca}\ \frac{\hat{r}_{12}^{trial\ 1}(t+1) - \hat{r}_{23}^{trial\ 2}(t+1)}{\left|\hat{r}_{12}^{trial\ 1}(t+1) - \hat{r}_{23}^{trial\ 2}(t+1)\right|} \tag{2.14}$$

$$\mathbf{r}_3^{trial\ 3}(t+1) = \mathbf{r}_3(t) + \delta_{ca}\ \hat{r}_{23}^{trial\ 2}(t+1)$$

The hard-sphere rearrangements illustrated by Equations (2.9) and (2.14), with their emphasis on the motion of small clusters, certainly look quite different from what one might expect to come from analogous calculations for continuous potentials, where one follows the full 3N-dimensional gradient of the system's potential energy.  On the other hand, even realistic intermolecular potentials have sharply varying repulsive cores.

We will examine the extent to which the formal hard-sphere/soft-sphere differences carry over to their respective geodesic pathways when we contrast the two in Sec. IV.

## III. MODELS AND COMPUTATIONAL DETAILS

***Models***:  The principal model we study in this paper is a crystallization-resistant binary mixture of different-sized hard spheres frequently used in the literature [15-17,23,24,29,61] and considered in detail by Flenner, Zhang, and Szamel.[16]  The system is a 50:50 mixture of small spheres (diameter $\sigma_s$) and slightly bigger spheres (diameter $\sigma_b$ = 1.4 $\sigma_s$).  The potential energy is therefore of the form of Eqs. (2.3) and (2.4) with species $\alpha$, $\beta$ = $b$ or $s$, and pair potentials given by

$$u_{\alpha\beta}(r) = \begin{cases} \infty & , \ 0 \leq r < \sigma_{\alpha\beta} \\ 0 & , \qquad r \geq \sigma_{\alpha\beta} \end{cases} \qquad , \qquad (3.1)$$

$$\sigma_{\alpha\beta} = \tfrac{1}{2}\left(\sigma_\alpha + \sigma_\beta\right)$$

The presence of two different sized spheres helps keep the system fluid over a wide range of densities, but as the packing fraction $\phi$ increases towards values close to the empirical mode coupling transition $\phi_c = 0.590$,

$$\phi = \frac{\pi\rho}{6}\left(x_b\sigma_b^3 + x_s\sigma_s^3\right) \qquad , \qquad \rho = \frac{N_b + N_s}{V} = \frac{N}{V} \qquad (3.2)$$

where, $N$ is the total number of particles, $V$ the volume, and for us, the mole fractions

$$x_b = N_b\big/\left(N_b + N_s\right) = \tfrac{1}{2} \qquad , \qquad x_s = N_s\big/\left(N_b + N_s\right) = \tfrac{1}{2} \, ,$$

the system slows dramatically and begins to exhibit many of the familiar characteristics of supercooled liquids.[16]

For comparison, we also briefly consider the single-component hard-sphere system ($\sigma_b = \sigma_s = \sigma$), and the Kob-Andersen Lennard-Jones binary mixture model [81] with species $\alpha$, $\beta = A$ or $B$, and pair potentials given by

$$u_{\alpha\beta}(r) = 4\varepsilon_{\alpha\beta}\left[\left(\sigma_{\alpha\beta}/r\right)^{12} - \left(\sigma_{\alpha\beta}/r\right)^{6}\right] \tag{3.3}$$

$$\varepsilon_{AA} = \varepsilon \quad, \quad \varepsilon_{BB} = 0.5\varepsilon \quad, \quad \varepsilon_{AB} = 1.5\varepsilon$$

$$\sigma_{AA} = \sigma \quad, \quad \sigma_{BB} = 0.88\sigma \quad, \quad \sigma_{AB} = 0.8\sigma$$

All of the Kob-Andersen calculations reported here use $N = 108$ particles ($N_A = 87$, $N_B = 21$).

***Geodesic calculations***: The hard-sphere geodesic calculations we report in this paper were carried out using the procedure described in Sec. II. We employed a simple downward-sloping linear form for our fictitious pair potential

$$v_{\alpha\beta}(x) = \sigma_{\alpha\beta}^2 \ (1-x) \tag{3.4}$$

We note, though, that trials with an alternate potential providing a continuous force at x = 0 and 1

$$v_{\alpha\beta}(x) = \sigma_{\alpha\beta}^3 \ \left(\frac{1}{3}x^3 - \frac{1}{2}x^2 + \frac{1}{6}\right)$$

showed no discernable differences in geodesic pathways. The step size parameters needed to implement Eqs. (2.1) and (2.8) were taken to be

$$\delta_{\text{total}} = 10^{-2} \ \sigma_s \quad, \quad \delta_{\text{ca}} = 10^{-4} \ \sigma_s$$

The requisite geodesic endpoints were computed via a standard hard-sphere molecular dynamics simulation (*vide infra*): Pairs of initial and final liquid configurations were derived by propagating the molecular dynamics from each selected initial configuration until the desired end-to-end separation $\Delta R$ was achieved, and then waiting another $10^5$ steps before identifying the next initial configuration. Results are typically averaged over geodesics derived with 5 such statistically independent end-point pairs.

Both as a test of the geodesic-finding algorithm, and to get as close as possible to true geodesic paths, we find it useful to refine our initially computed hard-sphere geodesics via local optimizations, much as we did with soft-potential geodesics.[64] Given a candidate geodesic path going from a configuration $\mathbf{R}(i)$ to a configuration $\mathbf{R}(f)$, we pick a configuration $\mathbf{R}$ along the path, displace it by a small amount to some new energetically-allowed configuration $\mathbf{R}'$, and find new geodesic candidates going from $\mathbf{R}'$ to $\mathbf{R}(i)$ and $\mathbf{R}'$ to $\mathbf{R}(f)$. If the sum of the two path lengths is less than the length of the original geodesic candidate, the (manifestly shorter) path $\mathbf{R}(i) \rightarrow \mathbf{R}' \rightarrow \mathbf{R}(f)$ is taken to be the new candidate for the $\mathbf{R}(i) \rightarrow \mathbf{R}(f)$ geodesic path. In the calculations presented in Sec. IV, we repeat this process until five successive $\mathbf{R}(i) \rightarrow \mathbf{R}' \rightarrow \mathbf{R}(f)$ candidates fail to achieve a lower total path length. With our procedure for generating random configurations $\mathbf{R}'$,[82] we typically find that this approach will shorten geodesic path lengths by amounts on the rough order of 10%, a relatively small correction that leads to differences in predicted diffusion constants difficult to see on the scale of the figures presented in this paper.

***Hard-sphere molecular dynamics and diffusion constants***: The equilibrated initial phase-space points needed to run molecular dynamics trajectories for the binary hard-sphere system were computed by first placing the big and small particles in different halves of an fcc lattice. For packing fractions $\phi \leq 0.50$, the system had a set of Gaussian distributed velocity vectors $\mathbf{v}_j$ assigned to each of the $N$ particles $j$, thereby defining the ratio

$$\frac{k_B T}{m} \equiv \frac{1}{3N} \sum_{j=1}^{N} \mathbf{v}_j^2 \qquad ,$$ (3.5)

and the system was then relaxed by allowing it to evolve for $10^7$ molecular dynamics steps. For larger packing fractions, the initial big/small-segregated lattice could itself have illegal particle overlaps, so the starting configurations were constructed by first taking our $\phi = 0.50$ fcc lattice and subjecting it to a mechanical contraction protocol.[83] The resulting configuration was then imbued with kinetic energy and relaxed in exactly the same fashion as with the lower packing fraction cases.

The diffusion constants we needed were calculated by integrating molecular dynamics-derived velocity autocorrelation functions. In particular, the separate big ($\alpha$=b) and small ($\alpha$=s)-particle diffusion constants $D_\alpha$ come from Simpson's rule evaluations of

$$D_\alpha = \frac{1}{3} \int_0^T dt \left\langle \frac{1}{N_\alpha} \sum_{j_\alpha=1}^{N_\alpha} \mathbf{v}_{j_\alpha}(0) \bullet \mathbf{v}_{j_\alpha}(t) \right\rangle$$ (3.6)

with the results checked for convergence with respect to changes in the integration intervals T and in the number of phase-space points sampled in those intervals. All of the diffusion constant results reported here are averages over 10 trajectories separated by $2 \times 10^5$ time steps. Diffusion constants for both big and small particles are reported in reduced units as

$$D^* \equiv D \frac{1}{\sigma_s} \sqrt{\frac{m}{k_B T}}$$ (3.7)

The molecular dynamics procedures themselves were checked by examining predicted radial distribution functions and velocity autocorrelation functions, but we note that comparison of our calculated diffusion constant for the single-component hard-sphere case with those in the literature yielded agreement within our error bars over a wide range of densities.[84]  Further calculational details are reported elsewhere.[85]

# IV. RESULTS

So, how different are the geodesic pathways through a hard-sphere potential energy landscape from the geodesics seen with softer potentials? The key idea we exploited with softer potentials is that increases in geodesic path lengths are one of the causes of slow dynamics. The decrease in landscape energy associated with decreasing the temperature puts increasingly stringent constraints on where paths can venture in the landscape – meaning that geodesics automatically become increasingly tortuous as $T$ goes down.[64] But, as we can see from Fig. 1, the same lengthening occurs for hard-sphere systems as the density increases. Raising the packing fraction of our binary hard-sphere fluid by a factor 10 triggers a hundred-fold growth in the geodesic path length. A decrease in (what is often called) the "free volume" [86] apparently translates rather specifically into increased inaccessibility of portions of the potential energy landscape.

Of course, for this path length behavior to provide any meaningful quantitative information about the landscape, our findings cannot depend on how the endpoints are chosen. That is, for each thermodynamic state, the lengths along the contour of the geodesic $g$ must themselves scale with the direct initial-($i$)-to-final-($f$) distance in configuration space $\Delta R$.[64,67] But,

$$\Delta R = \sqrt{\sum_{j=1}^{N} \left( \mathbf{r}_j(f) - \mathbf{r}_j(i) \right)^2} \qquad (4.1)$$

(with sum over all $N$ particles in the system). Hence, our requirement is that the ratio $g/\Delta R$ must be invariant with respect to either increases in the average per-particle displacement when $N$ is fixed, or increases in $N$ when the average per-particle displacement is fixed (which implies choosing $\Delta R \sim \sqrt{N}$ ). Figure 1 shows that both of these invariances are accurately preserved. Given that our connection between path lengths and diffusion constants, Eq. (1.2), is asymptotic in $\Delta R$, these invariance not only serve as a check that our geodesics are being computed correctly, they act as a useful confirmation that our simulations are not encountering finite-size effects.

How well, then, do geodesic path lengths predict the rapid decline in diffusion constants seen as we approach glassy behavior in hard-sphere fluids? The answer is given in Figs. 2 and 3. The single-component hard-sphere fluid system depicted in Fig. 2 is presented mainly for comparison: one can see that it has a moderate slowing down of its dynamics as the packing fraction ($\phi = (\pi/6) \rho\sigma^3$) heads towards the freezing point ($\phi = 0.49$) and that that slowing down is nicely predicted by purely landscape information (the trend in geodesic path lengths) – without requiring any dynamical input.

The main result of the paper, though, is shown in the analogous results for the hard-sphere mixture system, Fig. 3. The mixture does not crystallize, so it continues to slow as the packing fraction is increased beyond 0.5. In fact, this system displays a 4 order-of-magnitude decrease in diffusion constants over the same 10-fold increase in packing fraction we examined in Fig. 1. The remarkable feature is that this decline is accounted for *quantitatively* simply by the rate of growth of the geodesic path lengths.[87]

Aside from a single overall multiplicative constant needed to set the scale of the diffusion constants determined by Eq. (1.2), there has been no dynamical information used in constructing either Fig. 2 or Fig. 3. We did use molecular dynamics trajectories to provide a sample of initial and final configurations for our geodesic paths, but that was purely a matter of convenience; a Monte Carlo sampling scheme would have worked equally well.[64]    It therefore seems clear that understanding the specific potential surface features that define the geodesics in these systems should be sufficient for us to understand why the relaxation times behave the way that they do.

There should be other kinds of information embedded in the geodesics as well. To the extent that geodesic pathways define the inherent dynamics of our systems, examining them should also reveal the increasing tendency towards dynamical heterogeneity seen in the actual molecular dynamics.[15,88]    With perfect Fickian diffusion, for example, the probability density of the logarithm of the individual particle displacements

$$\log_{10} \, \delta r_j(t) = \log_{10} \, \left| \mathbf{r}_j(t) - \mathbf{r}_j(0) \right| \qquad , \qquad (j = 1, \ldots, N) \qquad (4.2)$$

can be shown to have a shape that is independent of time $t$,[89] but at times on the order of the $\alpha$ relaxation, the binary hard-sphere model has a distribution that broadens with increasing density, evidencing increasing level of dynamical diversity among the different particles.  When $\phi$ reaches 0.58, there are actually a clear signs of bimodality in the distribution.[16]

Our results in this paper only extend to $\phi = 0.57$,[90] but by examining the geodesics it is nonetheless possible to see how the beginnings of this heterogeneity stem from the underlying potential surface.  If we look at an analogue of Eq. (4.2), the distribution of the (logarithms of the) contour lengths along the entire geodesic paths taken by the individual small spheres $g_{js}$, scaled by $\Delta R_{js}$, the corresponding end-to-end distance travelled by each sphere,

$$\log_{10} \delta_{js} \equiv \log_{10} \left( g_{js} / \Delta R_{js} \right) \qquad , \qquad (j = 1, \ldots, N_s) \qquad , \tag{4.3}$$

the resulting probability densities show a clear growth in the root-mean-square width

$$\Delta = \left\langle \left( \log_{10} \delta_{js} - \left\langle \log_{10} \delta_{js} \right\rangle \right)^2 \right\rangle^{1/2} . \tag{4.4}$$

as the packing fraction increases (Fig. 4). There is no evidence, at least within our density range, of a second peak appearing in the distribution, but as density increases and geodesic pathways for the whole system becomes less and less direct, the shapes of the geodesics of individual particles evidently become progressively more diverse.[88]

A different kind of information contained in the geodesic is the average number of particles contributing significantly to the progress along the geodesic. As we noted in our soft-particle work,[66] it is possible to measure that average by for any given step along the pathway via the *participation ratio*, *n*. Writing the unit vector describing the direction travelled in the *3N*-dimensional configuration space during the step $\mathbf{R}(t) \to \mathbf{R}(t+1)$

$$\delta \hat{\mathbf{R}}(t) = \frac{\mathbf{R}(t+1) - \mathbf{R}(t)}{|\mathbf{R}(t+1) - \mathbf{R}(t)|} = \sum_{j\mu} c_{j\mu}(t) |j\mu\rangle = \sum_{j=1}^{N} \mathbf{c}_j(t) |j\rangle \qquad , \tag{4.5}$$

$$\sum_{j=1}^{N} \left| \mathbf{c}_j(t) \right|^2 = 1 \tag{4.6}$$

(where *j* labels the particles and $\mu = x, y, z$ labels the Cartesian coordinates), in terms of the associated (3-dimensional) single-particle components, $\mathbf{c}_j(t)$, allows us to construct the quantity

$$n(t) = 1 \Big/ \sum_{j=1}^{N} \left| \mathbf{c}_j(t) \right|^4 \tag{4.7}$$

which reports the effective number of particles participating in the step.

The rationale for this claim is the same as it was when this construction was used to assess the degree of localization of excitations in disordered systems:[91] if one ever did have exactly $m$ of the $N$ particles contributing, and all doing so equally, all of the $N$-$m$ non-participating particles $i$ would have $\left| \mathbf{c}_i \right| = 0$. Normalization, Eq. (4.6), would then imply that each of the remaining, participating, particles $k$ would have $\left| \mathbf{c}_k \right| = m^{-1/2}$, which would make $n = m$. In less idealized situations, $n$ is only a statistical measure rather than an exact count, but its properties are revealing. As with the excitation problem, our participation ratio ought to be independent of $N$ when our steps are localized and ought to scale with $N$ when the steps are delocalized.

The simplest portrait of the participation in our geodesics is the participation number averaged over the entire length of the geodesic.[66]

$$\langle n \rangle = \int_0^1 d\tau \; n(\tau) \tag{4.8}$$

We can get at this average from the (unevenly spaced) discretized representation of paths given by our calculations if we convert Eq. (4.8) into a sum over the discrete contour lengths $s(t)$ traversed at each step along the geodesic. In terms of the requisite $s(t)$ and the corresponding complete geodesic path length $g$,

$$s(t) = \left| \mathbf{R}(t+1) - \mathbf{R}(t) \right| \quad , \quad g = \sum_{t=0}^{M-1} s(t)$$

(with *M* the number of steps in the path), our working expression for the average participation number is thus

$$\langle n \rangle = \sum_{t=0}^{M-1} \left( \frac{s(t)}{g} \right) n(t) \quad .$$

The average participation numbers obtained for our binary hard-sphere system (Fig. 5) are clearly macroscopic (of order *N*) and do not vary strongly as the glass transition is approached, reflecting the fact that the net transition from initial to final configurations in our geodesics involves most of the system, regardless of the density. The particular numerical results shown are rather robust: neither recalculating them with N = 256 instead of 108, nor changing between optimized and unoptimized geodesics, (neither of which are shown here) yield any significant difference in the hard-sphere plots.[85] Carrying out the same calculation for the Kob-Andersen soft-sphere model [81] also reveals much the same kind of macroscopic values for the participation numbers in the geodesics of that system.

However there is an interesting difference in the trend as the hard- and soft-particle systems approach their respective glass transitions – a difference that is indicative of a much more significant disparity between the geodesic pathways of the two. Suppose we look not at the average participation numbers but at the probability distribution of the individual-step participation numbers *n(t)* (Figs. 6 and 7, top panels). The contrast in those results seems to indicate that the natures of the geodesics pathways in hard and soft systems have to be fundamentally distinct.

One way to understand this distinction is to note that there is a relationship between what is often called the non-Gaussian parameter [16,27,30,31,33,88,89,92] $\alpha_2$ and the participation ratio $n$. If the single-particle displacements defined in Eq. (4.2), $\delta r(t) = |\delta \mathbf{r}(t)|$, had a perfect 3-dimensional Gaussian distribution at any time $t$, it is easy to show that

$$\alpha_2(t) = \frac{3}{5} \frac{\langle \delta r^4(t) \rangle}{\langle \delta r^2(t) \rangle^2} - 1 \tag{4.9}$$

would be identically zero, so the magnitude of this quantity is often used to measure how non-Gaussian the displacement distribution is in supercooled liquids. But, the participation ratio prescribed by Eqs. (4.5)-(4.7) looks at just this quantity for single steps along the geodesic. Defining the net single-step displacement of the system to be $\Delta R(t) = |\mathbf{R}(t+1) - \mathbf{R}(t)|$ means that we can write the component vectors

$$\mathbf{c}_j(t) = \delta \mathbf{r}_j(t) / \Delta R(t) \quad , \quad \Delta R^2(t) = \sum_{j=1}^{N} \left| \delta \mathbf{r}_j(t) \right|^2$$

If we then interpret our averages as averages over the different particles, the participation ratio, Eq. (4.7), becomes

$$n(t) = N \frac{\left( \frac{1}{N} \sum_{j=1}^{N} \left| \delta \mathbf{r}_j(t) \right|^2 \right)^2}{\frac{1}{N} \sum_{j=1}^{N} \left| \delta \mathbf{r}_j(t) \right|^4} = N \frac{\left\langle \left| \delta \mathbf{r}(t) \right|^2 \right\rangle^2}{\left\langle \left| \delta \mathbf{r}(t) \right|^4 \right\rangle}$$

meaning that,

$$\frac{n(t)}{N} = \frac{3}{5} \frac{1}{1 + \alpha_2(t)} \quad , \tag{4.10}$$

so the participation ratio and non-Gaussian parameter carry *exactly* the same information.

Equation (4.10) has a number of immediate implications. For systems whose displacements are nearly Gaussian ($\alpha_2 = 0$), we can see that the participation ratio should be extensive with $n/N \approx 0.6$. Indeed (to within numerical error) that value is precisely what is found in geodesic studies of normal (non-supercooled) liquids at equilibrium.[66,93] However, if there were ever few-particle contributions to the dynamics ($n = 1, 2, 3, \ldots$), they would lead to extensive values of $\alpha_2$ (corresponding to large, positive values in any finite-$N$ simulation). Conversely, if we ever had cooperative dynamics involving more than the expected Gaussian numbers of particles ($0.6\,N < n \leq N$), we would see small negative values of $\alpha_2$ ($-0.4 \leq \alpha_2 < 0$).

By looking simultaneously at the participation-number and the non-Gaussian parameter distributions on the geodesic, the basic differences between the intrinsic dynamics of hard- and soft-sphere systems becomes clear. The soft, Kob-Andersen, system (Fig. 7), shows only extensive participation numbers with $n/N$ values peaked close to the Gaussian value at high temperature. As the system supercools, it becomes less Gaussian ($n/N$ values decrease from 0.6, so that small positive $\alpha_2$ values begin to appear, $0 < \alpha_2 < 2$), but there is no evidence of few-particle dynamics. The binary-hard-sphere system (Fig. 6), on the other hand, displays sharp peaks at large $\alpha_2$ that can unmistakably be traced to $n = 2, 3, 4, \ldots$ particle moves.[94]

The weights of these events actually decrease as the density increases, but they remain a consistent presence at all the densities. Their existence for hard spheres and absence for soft particles is, in fact, completely consistent with the far greater locality of the inter-particle forces for hard spheres. Given that, it may seem somewhat surprising that the hard sphere system also differs from the softer system in displaying supra-Gaussian collective moves. These moves become increasingly important and involve larger and larger fractions of hard-sphere systems as the density increases. But this distinction, as well, can be traced to the differences in potential energy landscapes. Soft spheres can apparently avoid high potential energy locations without requiring whole-system rearrangements. Hard spheres do not have this luxury.

# V. CONCLUDING REMARKS

For realistic intermolecular potentials, there are two separate ingredients in a landscape geodesic perspective on slow dynamics. One is the central tenet: that dynamics becomes slow when even the most efficient pathways through the potential-energy landscape become lengthy.[64] The other is that the "most efficient" can be investigated within the potential-energy-landscape ensemble,[63] which effectively posits that barrier hopping does not contribute significantly to the traversal of the landscape under certain conditions. It is the combination of the two ideas that implies that the most efficient paths are the shortest routes that never exceed a certain potential energy.

We had shown previously that this combination offers a useful way of understanding at least the onset of slow dynamics in fragile liquids.[64] However in turning in this paper to hard sphere systems, systems in which the barrier hopping is rigorously absent, we leave ourselves solely with the question of whether the non-barrier-hopping origins of slowing down are accounted for *precisely* by a concomitant growth in the lengths of the geodesic paths through the potential energy landscape. The answer is evidently yes. The evolution of this topological feature of the hard-sphere potential surface with density is apparently all we need to explain the precipitous drop in diffusion constants seen as these systems get progressively glassier.

The fact that the same kind of rapid geodesic lengthening occurs in non-hard-sphere examples makes it natural to wonder how much of the universal phenomenology of supercooled liquids can be traced directly to the behavior of the analogous hard-sphere problem. Because geodesics can be thought of as defining the "inherent dynamics," and because we know the specific sequence of molecular rearrangements associated with them, we should be able to make facile comparisons. Distributions of the lengths of the geodesics of individual hard spheres, for example, seem to reflect the non-dynamical origins of dynamical heterogeneity, and the differences between rotational and translational components of molecular geodesics [67] may lie at the heart of the frequently observed divergences between the relaxation rates of those two kinds of motions.[95]

But that same kind of examination is already showing that there are fundamental differences in detail between the geodesics of hard and soft systems. The potential surface of a true hard-sphere fluid is far more sensitive to local rearrangements than any realistic liquid would ever be. As a result, our look at the distribution of the geodesic equivalent of the non-Gaussian parameter – the participation ratio – shows that few-particle moves are much more prominent with hard spheres. Moreover, the inability of a hard-sphere potential to redistribute small irregularities in the potential energy density forces the system into a lengthy series of adaptations to particle moves by yet other particle moves, resulting in more many-particle moves as well.

Still, the basic success of the landscape geodesic viewpoint for hard spheres suggests that we should reconsider some of the views on potential energy landscapes expressed in the literature. The prevailing notion of travel through a landscape [53] as being a succession of transitions between basins (or traps),[96] and possibly, between meta-basins,[54] is obviously very different from the geodesic picture's take on landscapes – where the focus is on paths instead of points. From a geodesic theory standpoint, the ability that the inherent structure concept gives us to map progress along a trajectory rigorously and unambiguously into a series of basins is no guarantee that those basins have a consistent role in the dynamics. However, there is more to a landscape than its stationary points. It may very well be, for example, that the landscape border regions emphasized by Kurchan and Laloux [97] and by Keyes, Chowdhary, and Kim, [98,99] are important not because of their associated saddle points, but because geodesic paths tend to lie near them.

One critique frequently voiced about the entire landscape philosophy in the context of glasses is that landscapes seem better suited to explaining static structure than dynamics. How, one might wonder, can any information about the shape of the landscape help us understand why dynamic length scales are more crucial than static length scales in determining glassy dynamics?[17,18] The answer again lies in the ability to consider paths and not just points. Within a configuration-space-path language, static length scales are statements about which individual configuration space points are most likely; dynamic length scales are the corresponding statements about the correlation between these points found along the principal paths. The key from our perspective is the realization that "principal path" means geodesic, so that the geometry of the landscape also provides ready insights into the paths.

A particularly pointed version of this concern about whether potential energy landscapes embody dynamical considerations in a useful way is that of the facilitated-kinetics view.[100,101] The thought here is that what matters are not special locations on the 3N-dimensional potential surface (special many-particle arrangements), but the local dynamical rules that govern the motion of coarse-grained densities from place to place in three dimensions. This approach, which focuses attention on the precisely choreographed sequence of events necessary for motion in a supercooled liquid, regards these sequences (the "space-time histories") as the key elements.[102]  Advocates of this view have even sometimes referred to this perspective as being explicitly "non-topographic," presumably to emphasize how peripheral they believe the potential surface minima and saddle points are to understanding the universal features of the dynamics (despite the central roles of the former in determining equilibrium structure and thermodynamics).

But, again, a landscape analysis does not have to revolve around stationary points. Indeed, the geodesic version of landscape theory has a number of perspectives in common with the facilitated kinetics view.  Since *all* of the slow dynamics predicted by geodesic theory occurs because of the scarcity of facile pathways – that is, because of entropy barriers – geodesic theory agrees with the facilitated kinetics predictions that activated dynamics should occur above, as well as below, the mode-coupling temperature.[96,103]  The existence, or lack of existence, of potential energy surface saddle points under these conditions is irrelevant.[104]  More strikingly, pivotal roles for local dynamical constraints appears prominently in both theories: Whitelam and Garrahan [101] pointed out that the solutions to the Fokker-Planck equation in "real" space can be thought of as geodesics whose glassy character stems from the constraints that the dynamical rules impose.  In much the same way, geodesics through configuration space exhibit glassy behavior because of the constraint created by the (apparently nonlocal) landscape energy condition that $V(\mathbf{R}) \leq E_L$.  But in the case of hard spheres, that translates into a requirement that, at each step $\tau$ along the path, the distances between every pair of particles $j$ and $k$ have to obey the local, real-space (nonholonomic) constraints that $r_{jk}(\tau) \geq \sigma_{jk}$.

Geodesic paths are undeniably "topographic" in origin, but they may also be viewed as summaries of – and explanations for – the facilitation rules. And while it is possible to generate much of the phenomenology of glassy behavior by postulating rules that have no connection whatsoever to the system's potential energy (beyond that implied by detailed balance),[105] the rules in physical systems *have to* stem from the underlying potential surface. The question for us is whether the particular properties of the landscape responsible for defining the geodesics are as germane to the dynamics of soft particles as they apparently are for hard spheres.

## ACKNOWLEDGEMENT

This work was supported by NSF grant CHE-1265798.

dynamics is slow in such cases because the system needs to locate difficult-to-find channels in the energy landscape. See Refs. [12, 31, 71-75].

0.1, 0.1) $\sigma_s$. Displacements that create particle overlaps are rejected and the procedure for creating displaced configurations repeated until it produces a configuration with no overlaps. Whenever a particle move is accepted but leads to a net displacement | **R**´- **R** | < 0.1 $\sigma_s$, an additional random choice is made for a sphere to be displaced and the procedure is continued until the net displacement does reach at least 0.1 $\sigma_s$.

[87] The level of theory represented by Eq. (1.2) connects dynamics to the overall geodesic path length for the entire system. It therefore predicts that the self-diffusion constants of the big and small particles should have identical density dependences, which they largely do. Capturing the slight differences in dependences that are present would require further analysis of the contributions of the different components of the geodesics to the self and mutual diffusion constants.

[88] A Widmer-Cooper, P. Harrowell, and H. Fynewever, Phys. Rev. Lett. 93, 135701 (2004).

[89] E. Flenner and G. Szamel, Phys. Rev. E **72**, 031508 (2005).

[90] Our current techniques for calculating geodesics seem to work for even higher densities, but we found that our statistics at these densities were insufficient to merit publication. As we discovered in our previous study of the Kob-Andersen soft-sphere system (Ref. [64]), the principle computational difficulty encountered as the system becomes increasingly glassy lies *not* in finding geodesic between prescribed endpoints, but in locating connectable points in configuration space that are sufficiently separated to serve as geodesic endpoints. We assure connectability in our calculations by using molecular dynamics to generate our endpoints; that guarantees us that there is at least one path between them (albeit one far longer than that of a geodesic). Our inability to find *any* well-separated, connectable, pairs of configurations at densities beyond the empirical mode coupling transition, $\phi = 0.59$ (Ref. [16]) (along with our parallel inability to find such pairs for the Kob-Andersen model below its empirical transition temperature, Ref. [64]) makes us suspect that the empirical mode-coupling transition, despite its well-

known lack of relevance to the true laboratory glass transition (Ref. [8]), coincides with a genuine potential-energy-landscape percolation transition.

**Figure captions**

**FIG. 1.** (color online). The increase in geodesic path length $g$ with packing fraction $\phi$ for our binary-hard-sphere mixture model. For a given packing fraction, the geodesic path length scales with $\Delta R$, the net distance traversed by the path, as shown by the invariance of the ratio $(\Delta R/g)^2$ to (a) changes in $\Delta R$ while keeping the numbers of particles $N$ fixed (*upper panel*) and (b) changes in $N$ while keeping the net distance-travelled-per-particle $(\Delta R/\sqrt{N})$ fixed (*lower panel*).

**FIG. 2**. (color online). The dependence of the (reduced) diffusion constant $D^*$ on packing fraction $\phi$ for a single-component hard-sphere fluid. The figure compares the geodesic prediction from Eq. (1.2), which evaluates the diffusion constants (to within an overall multiplicative constant) solely from the geometry of the potential energy landscape, and the results from full molecular dynamics (MD) calculations. The geodesic results presented are averages over 5 optimized paths with $N = 108$ and $\Delta R = 108 \, \sigma_s$. Molecular dynamics results were also obtained by averaging over 5 trajectories of $N = 108$ sphere systems.

**FIG. 3**. (color online). The dependence of big-particle and small-particle (reduced) diffusion constants $D^*$ on packing fraction $\phi$ for a binary hard-sphere fluid. As in Fig. 2, the figure compares potential-energy-landscape-based predictions (*geodesic*) with results from full molecular dynamics (*MD*). The geodesic results presented are for the entire

(big plus small-particle) system and are thus the same in both panels except for different choices of an overall multiplicative constant. The geodesic curves represent averages over 5 paths with $N = 108$ and $\Delta R = 108\ \sigma_s$ using optimized paths for all but $\phi = 0.56$, 0.57. Molecular dynamics results were obtained by averaging over 10 trajectories of $N = 108$ particle systems.

**FIG. 4**. (color online). A landscape perspective on dynamical heterogeneity of a binary-hard-sphere fluid: the probability distribution of single-particle components of the geodesic path lengths. The upper panel looks at how the distribution, over different particles $j$, of (the logarithm of) the ratios of geodesic path lengths to the direct distances ($g_{js}/\Delta R_{js}$) changes with increasing packing fraction $\phi$ (with the packing fraction increasing from left to right). The lower panel plots the evolution of the corresponding root-mean square widths $\Delta$. All of these results were derived from geodesics with a total of $N = 108$ particles traversing a small-particle net distance $\Delta R_s = 90\ \sigma_s$, and averaged over 50 unoptimized paths.

**FIG. 5**. (color online). Average participation ratios for the geodesics of binary hard spheres and the Kob-Andersen model plotted as a function of the packing fraction $\phi$ and reduced temperature $k_B T/\varepsilon_{AA}$, respectively. To facilitate comparison, the scales for both models are arranged so that dynamics becomes progressively slower as you proceed from left to right, and the scales are translated so that the empirical mode-coupling transitions for both models occur at the same point (indicated by a vertical dashed line). Geodesic

pathways for both systems are computed with a total of $N = 108$ particles travelling a net distance $\Delta R = 108\ \sigma_s$ and the reported participation ratios averaged over 5 optimized paths.

**FIG. 6.** (color online). The distribution of particle numbers involved in individual steps along the geodesic pathways for *binary hard spheres* at a variety of different packing fractions. Shown in the upper panel are probability distributions of single-step participation fractions $n/N$, with the characteristic Gaussian value of this fraction indicated by a dashed vertical line. The lower panel displays the corresponding distributions of single-step non-Gaussian parameters $\alpha_2$ (with the value 0 denoting perfect Gaussian behavior). The discrete peaks at large $\alpha_2$ values stem from the indicated discrete numbers $n$ of participating particles. (For our simulation, Eq. (4.10) implies that $n = 2, 3, 4, \ldots$ should correspond to $\alpha_2 = 31.4, 20.6, 15.2, \ldots$ .) Geodesic pathways are computed with $N = 108$ and $\Delta R = 108\ \sigma_s$, and the particle-number distributions averaged over 20 unoptimized paths.

**FIG. 7.** (color online). The distribution of particle numbers involved in individual steps along the geodesic pathways for the *Kob-Andersen model* at a variety of different reduced temperatures. Other than the switch from hard spheres to a soft particle system, the computational details and details of the figure presentation are identical to those in Fig. 6. Curves in the upper panel correspond to progressively higher reduced temperatures

$k_B T/\varepsilon_{AA}$ as one proceeds from left to right; curves in lower panel (though barely distinguishable on this scale) have the opposite order.

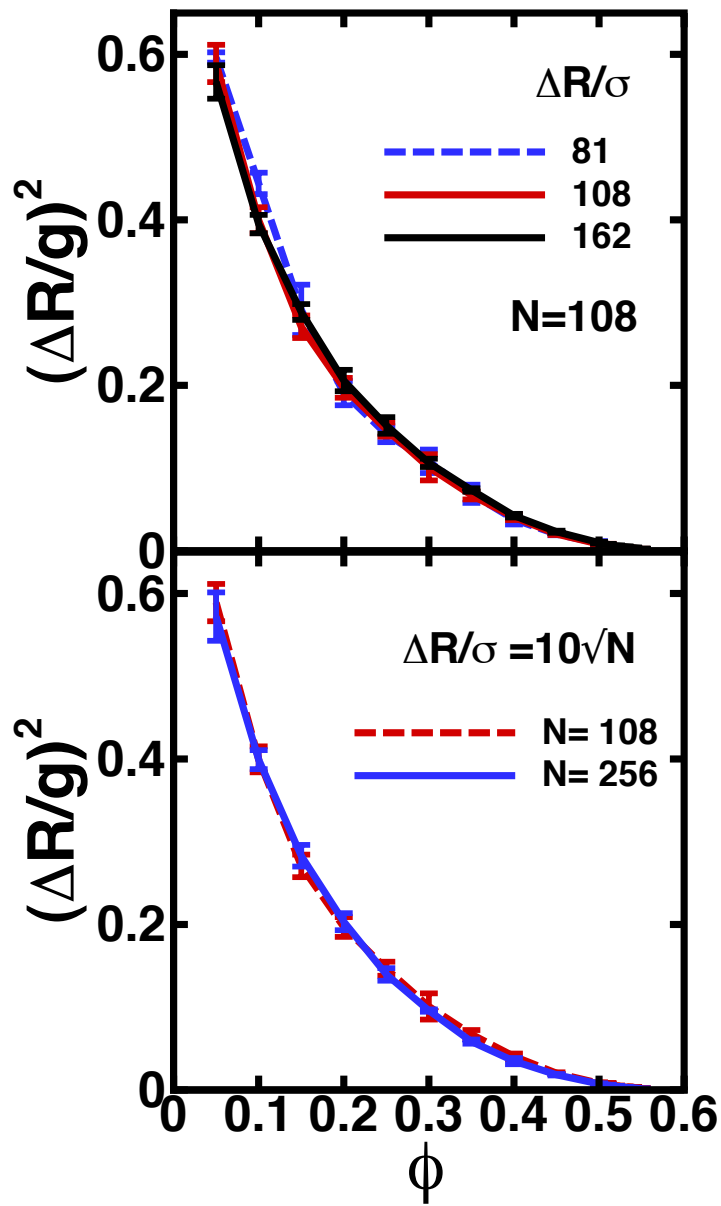

Figure 1

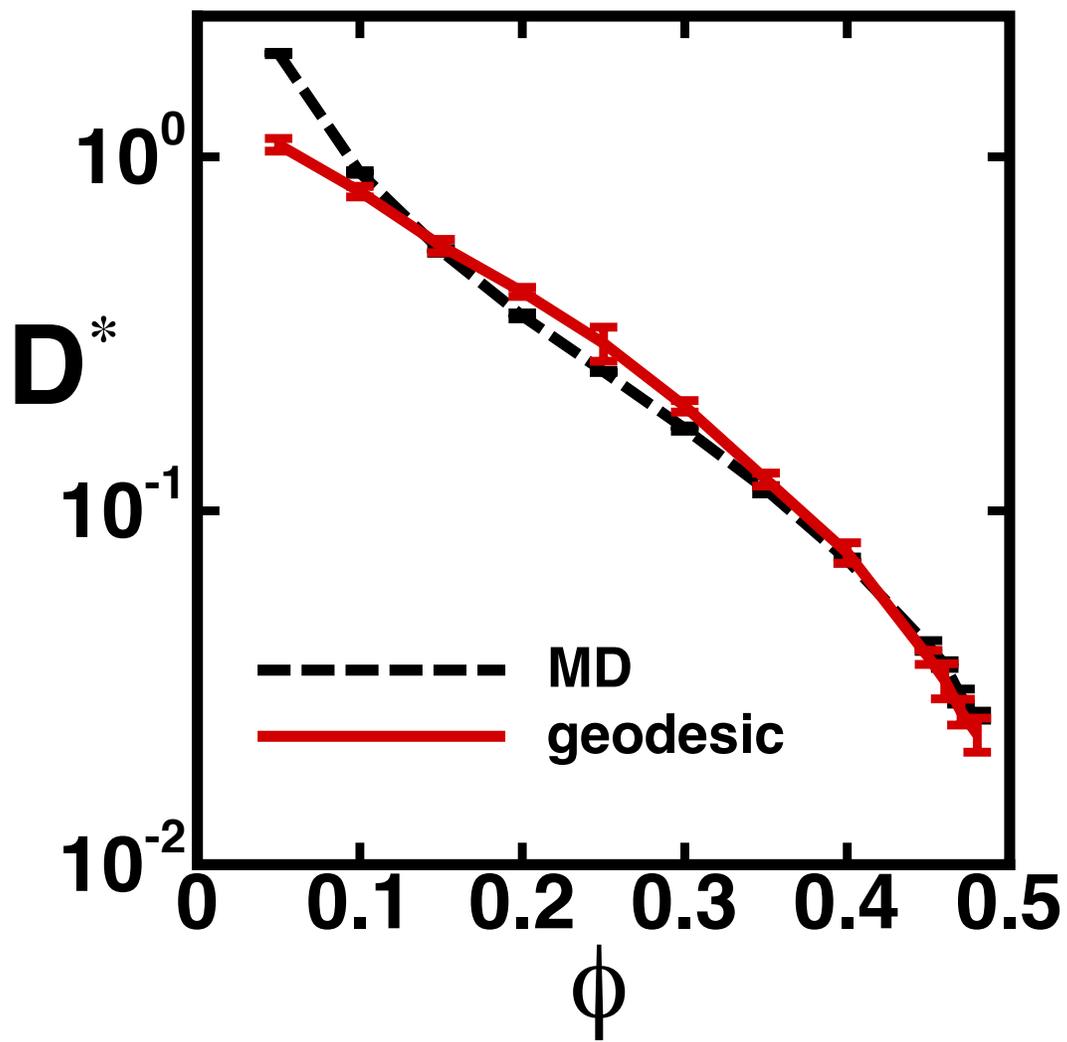

Figure 2

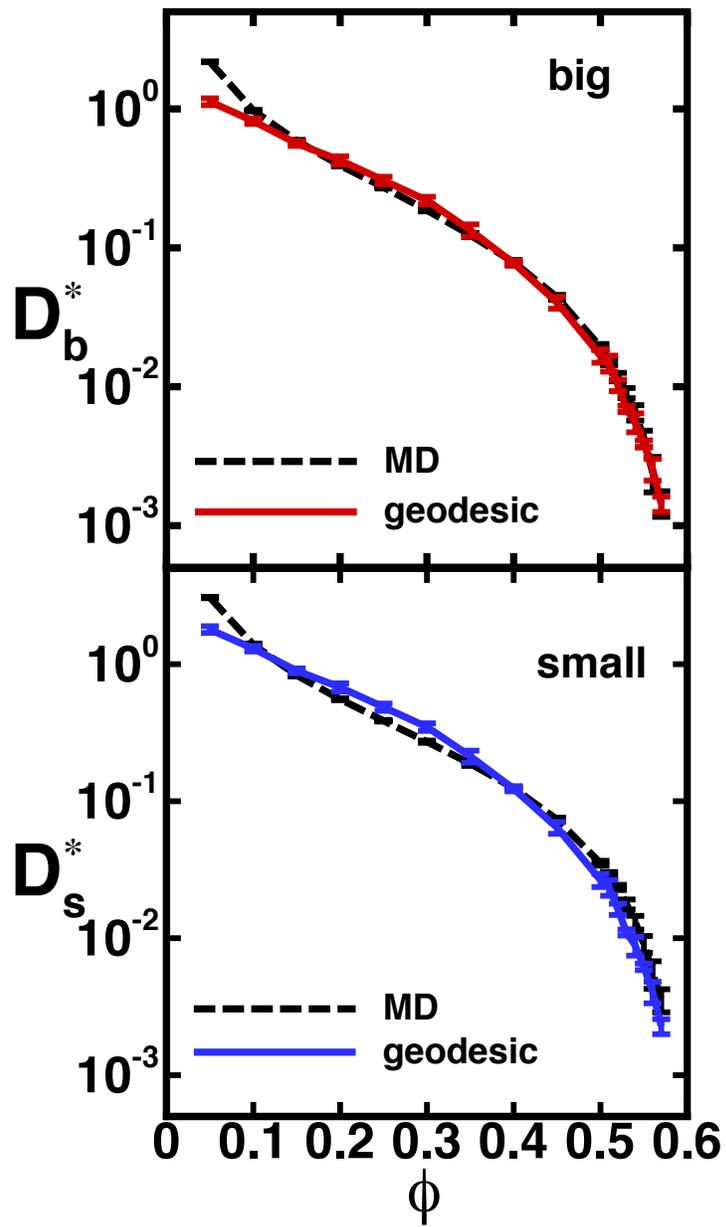

Figure 3

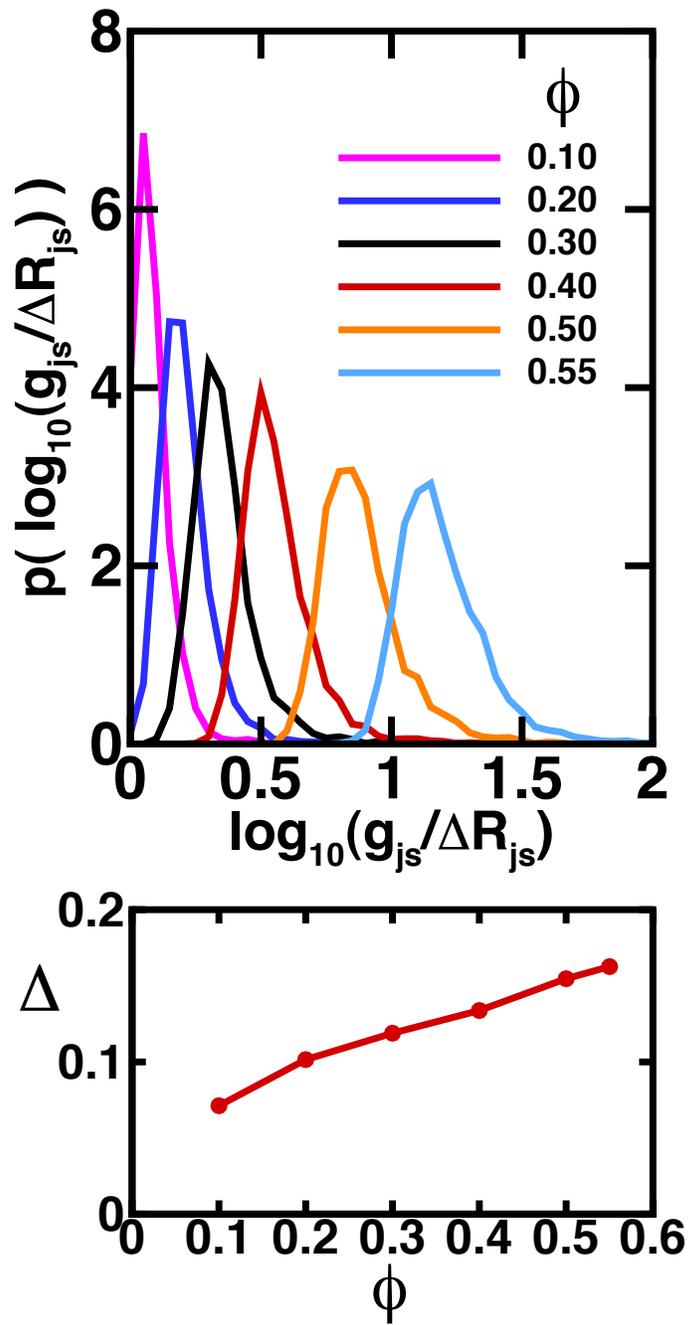

Figure 4

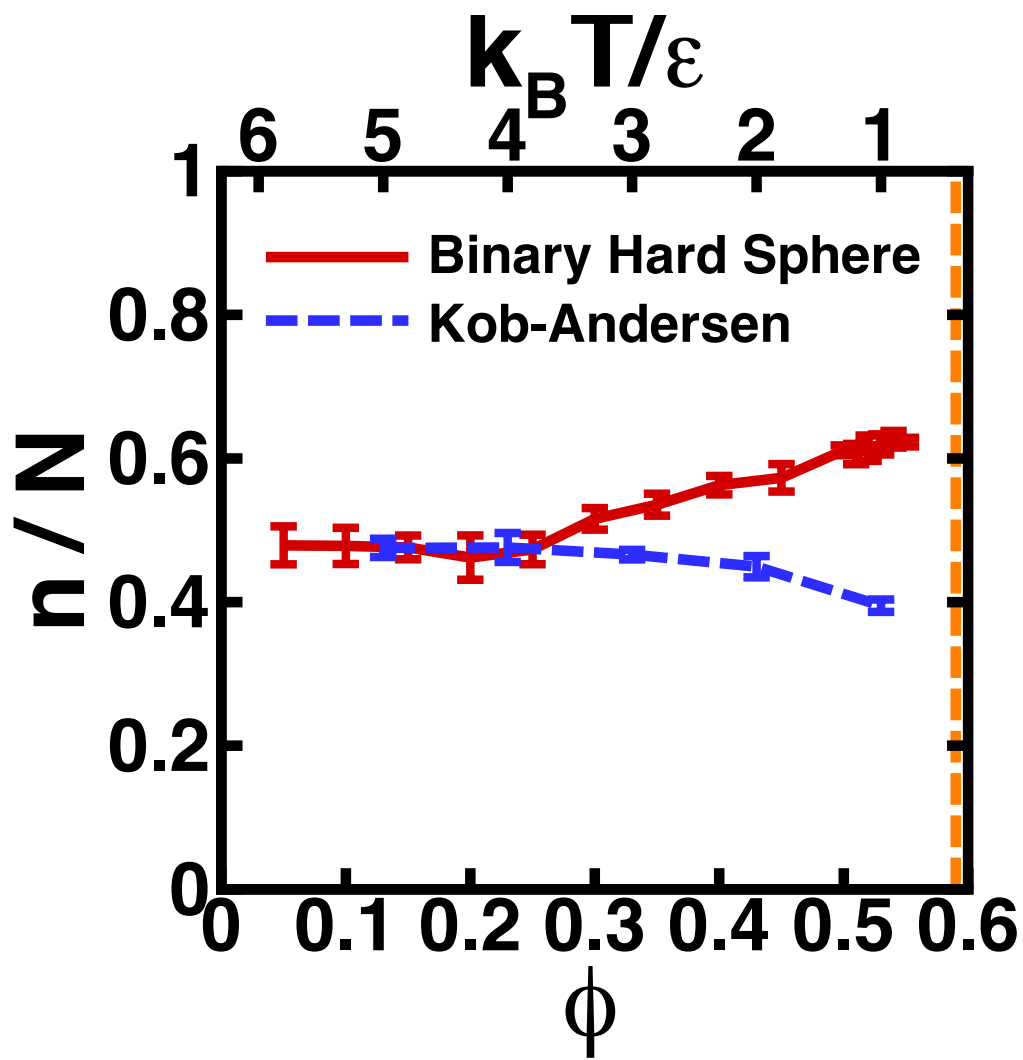

Figure 5

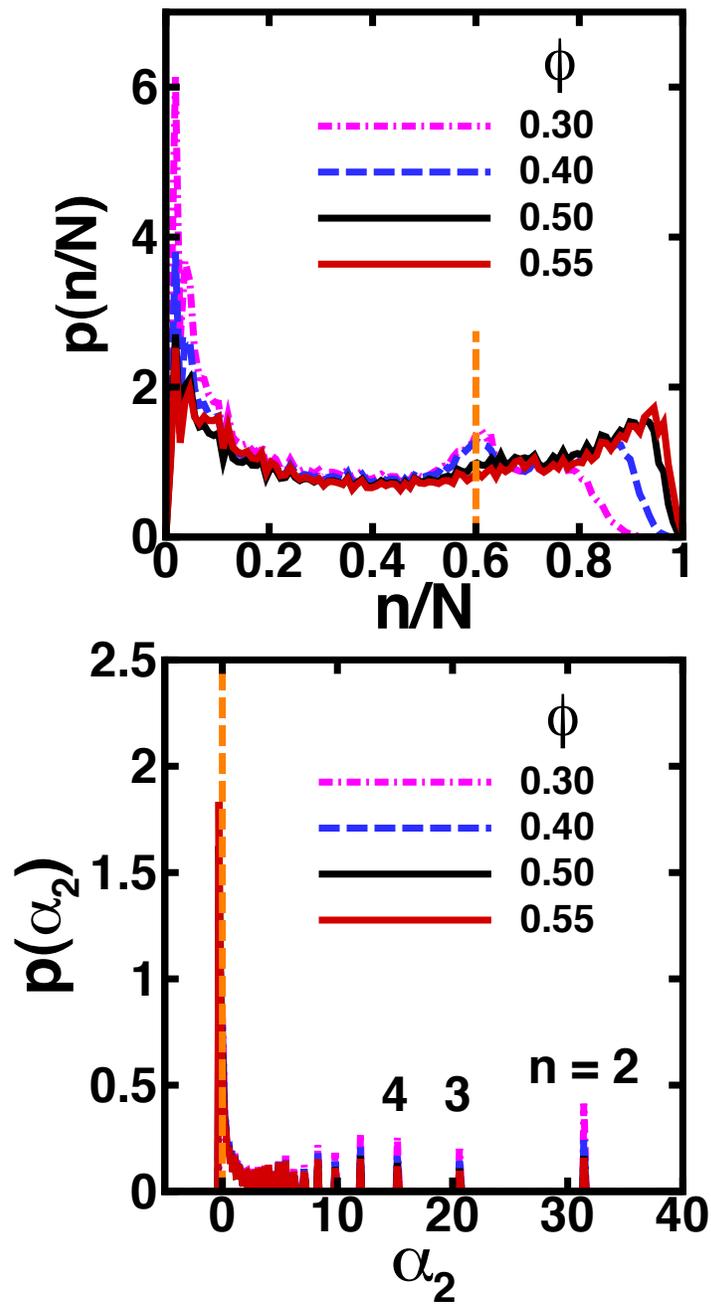

Figure 6

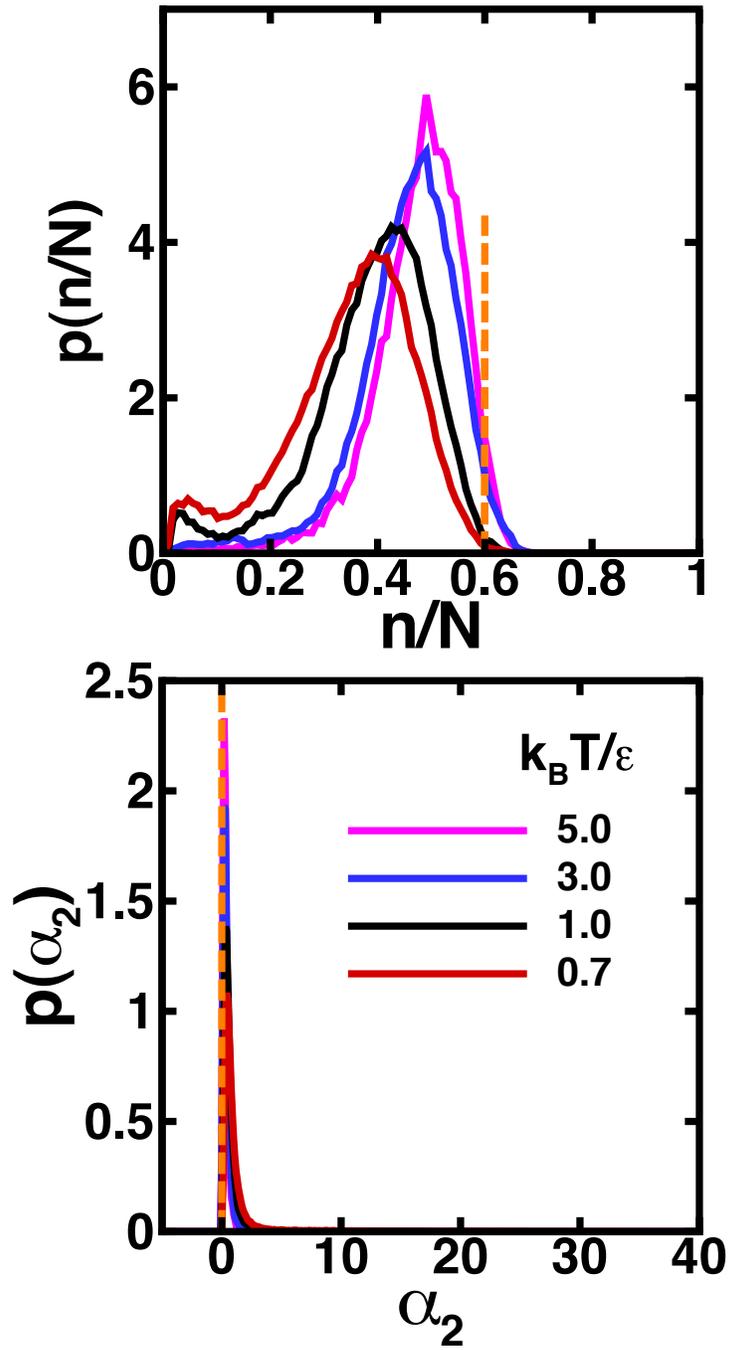

Figure 7